\pdfoutput=1
\documentclass[aps,showpacs,prl,twocolumn]{revtex4-1}
\usepackage{graphicx}
\usepackage{bm}

\begin{document}
\title{Reverse epitaxy of Ge: ordered and facetted  surface patterns} 
\author{Xin Ou}
\affiliation{Institute of Ion Beam Physics and Materials Research, Helmholtz-Zentrum Dresden-Rossendorf, Bautzner Landstr. 400, 01328 Dresden, Germany}
\author{Adrian Keller}
\affiliation{Institute of Ion Beam Physics and Materials Research, Helmholtz-Zentrum Dresden-Rossendorf, Bautzner Landstr. 400, 01328 Dresden, Germany}
\author{Manfred Helm}
\affiliation{Institute of Ion Beam Physics and Materials Research, Helmholtz-Zentrum Dresden-Rossendorf, Bautzner Landstr. 400, 01328 Dresden, Germany}
\author{J\"urgen Fassbender}
\affiliation{Institute of Ion Beam Physics and Materials Research, Helmholtz-Zentrum Dresden-Rossendorf, Bautzner Landstr. 400, 01328 Dresden, Germany}
\author{Stefan Facsko}
\email{s.facsko@hzdr.de}
\affiliation{Institute of Ion Beam Physics and Materials Research, Helmholtz-Zentrum Dresden-Rossendorf, Bautzner Landstr. 400, 01328 Dresden, Germany}

\date{\today}

\begin{abstract}
Normal incidence ion irradiation at elevated temperatures, when  amorphization is prevented, induces novel nanoscale patterns of crystalline structures on elemental semiconductors by a reverse epitaxial growth mechanism: on Ge surfaces irradiation at temperatures above the recrystallization temperature of $250^\circ \mathrm{C}$ leads to self-organized patterns of inverse pyramids. Checkerboard patterns with fourfold symmetry evolve on the Ge (100) surface, whereas on the Ge (111) surface, isotropic patterns with a sixfold symmetry emerge. After high fluence irradiations these patterns exhibit well developed facets. A deterministic nonlinear continuum equation accounting for the effective surface currents due to an Ehrlich-Schwoebel barrier for diffusing vacancies reproduces remarkably well our experimental observations.
 \end{abstract}

\pacs{68.35.Ct, 79.20.Rf, 81.16.Rf, 81.65.Cf}
\maketitle 


Self-organized pattern formation in systems far from equilibrium is a fundamentally interesting phenomenon governed by the interplay of kinetic and diffusive mechanisms. In these systems the dynamics of pattern formation obeys universal laws with general scaling relations \cite{Barabasi:ti}. In addition, surface patterns with nanoscale dimensions are of technological interest for applications in sublithographic surface templating and for quantum dot device fabrication \cite{Teichert:2002uj}. They may be generated on surfaces by homoepitaxy \cite{VanNostrand:1995wn}, heteroepitaxy \cite{Chen:2012ii} or by energetic ion irradiation \cite{Facsko:1999vp,Valbusa:2002tw,Chan:2007df}. 
However, semiconductor surfaces become amorphous during conventional ion irradiation at room temperature. At these conditions periodic ripple patterns oriented perpendicular or parallel to the ion beam direction and isotropic, hexagonally ordered, dot or hole patterns have been observed \cite{Chan:2007df} and are independent of the crystal structure \cite{Facsko:2002tt}. The origin of these patterns is attributed to an interplay of a surface instability due to sputtering and mass redistribution together with surface relaxation mechanisms \cite{Bradley:1988uf,Davidovitch:2007cl}. Below incidence angles of 50$^\circ$ smoothing is expected to dominate on amorphized elemental materials \cite{Madi:2011iy}. On the other hand, metals and metal oxide surfaces remain crystalline during ion irradiation at room temperature \cite{Valbusa:2002tw}. They exhibit a much higher complexity of pattern formation due to additional instabilities resulting from anisotropies in surface diffusion and due to biased diffusion across step edges \cite{Rusponi:1998wc}. Strong similarities with homoepitaxy have been identified: layer by layer erosion has been observed on metals \cite{Poelsema:1984eq} as well as on semiconductor surfaces \cite{Bedrossian:1991wh}. Furthermore, similar to mound formation in epitaxy, pit formation has been observed experimentally on ion irradiated metal surfaces \cite{Michely:1991wm,Stroscio:1995ce}. Although the formation of pits has also been seen in low fluence irradiations of semiconductors \cite{Chey:1995te,Kim:2003bd,Zandvliet:1997hg}, dense and ordered patterns of facetted nanostructures, as found in homoepitaxy and heteroepitaxy, have not been observed until now on ion irradiated semiconductor surfaces. 

In this letter we present the formation of regular patterns of crystalline structures induced by normal incidence ion irradiation of an elemental semiconductor, Ge, at elevated temperatures. At temperatures above the recrystallization temperature of $250^\circ \mathrm{C}$, ion induced bulk defects are dynamically annealed and the surface remains crystalline. By only increasing the irradiation temperature, instead of inducing surface smoothing, novel checkerboard patterns with crystalline facets appear. These patterns exhibit the symmetry of the crystalline surface, i.e.\ on the (100) surface the structures are oriented along the $\langle100\rangle$ direction whereas on the Ge (111) surface pit patterns with a six fold symmetry develop. They strongly resemble mound patterns in homoepitaxial and heteroepitaxial growth \cite{VanNostrand:1995wn}, but are reversed. The mechanism can thus be interpreted as Òreverse epitaxyÓ.  In analogy with the Villain instability in homoepitaxy resulting from the Ehrlich-Schwoebel (ES) barrier for an adatom descending a monoatomic step \cite{Villain:1991iw,Amar:1996us} we conclude that the formation of ion induced crystalline patterns results from the existence of an ES barrier for the ascending of surface vacancies created by sputtering. Based on the proposed atomistic mechanisms we derived a continuum equation for reverse epitaxy which describes remarkably well the experimentally observed surface evolution. 


Ge (100) and Ge (111) samples cut from epi-ready Ge wafers (root mean square roughness $w_{rms} = 0.7\,\mathrm{nm}$) were irradiated by a broad 1 keV Ar$^+$ ion beam at normal incidence without any pre-treatment. The irradiations were performed in a high vacuum chamber (base pressure $~10^{-8}$ mbar) equipped with a Kaufman ion source with a single graphite grid extraction with 5 cm diameter. During irradiation the chamber is backfilled with Ar at $10^{-4}\,\mathrm{mbar}$. In order to avoid metallic contaminations from the sample holder the samples with a typical size of $5\times5\,\mathrm{mm}^2$ were glued to a Si plate with size of $10\times15\,\mathrm{mm}^2$. The samples were heated by a boron nitride heater from the backside of the Si plate. The surface temperature was monitored during ion irradiation by a pyrometer operated in the wavelength range of 2-2.8 $\mathrm{\mu m}$ that was calibrated externally by a thermocouple. The surface topography was analyzed after irradiations ex-situ by a multimode atomic force microscope (AFM) from Veeco in tapping mode.

\begin{figure}
\includegraphics[width=\columnwidth]{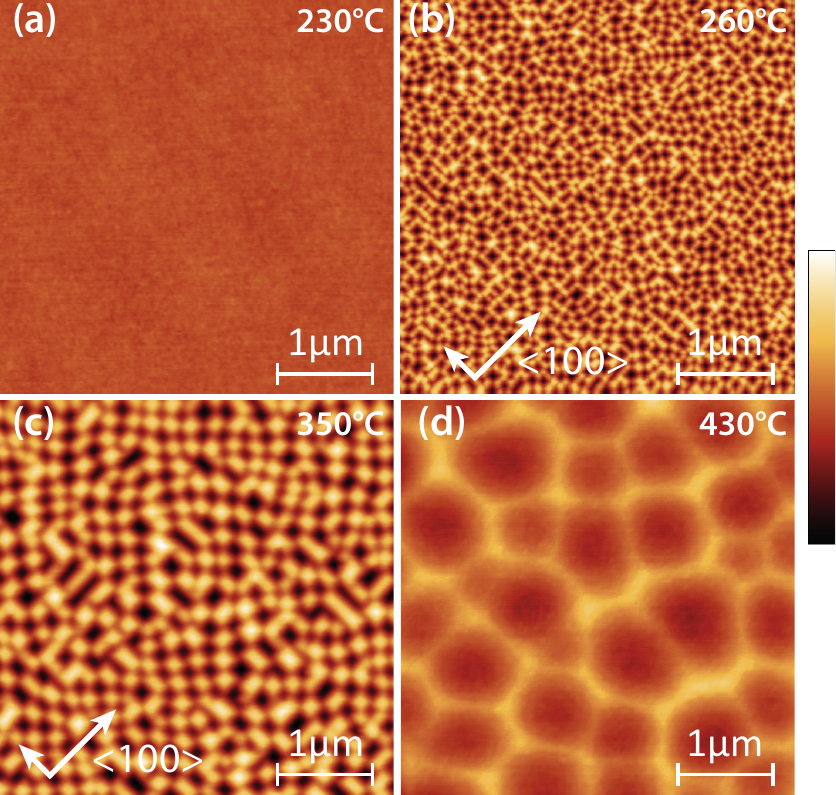} 
\caption{\label{AFMTemp}AFM height images ($4\times 4\ \mu m^{2}$) of Ge (100) surfaces after ion irradiation with an ion fluence of $3\times10^{18}\,\mathrm{cm}^{-2}$ at temperatures of $230^\circ$C, $260^\circ$C, $350^\circ$C, and $430^\circ$C, respectively. The $\langle100\rangle$ crystal directions are marked by arrows in (b) and (c). The height color bar on the right side represents 8 nm for (a) and (d), and 38 nm for (b), (c), respectively.
}

\end{figure}

In Fig.\ \ref{AFMTemp} AFM images of Ge (100) surfaces irradiated with ion fluences of $3\times 10^{18}\,\mathrm{cm}^{-2}$ and different surface temperatures ranging from $230^\circ \mathrm{C}$ to $430^\circ \mathrm{C}$ are shown. At temperatures below $250^\circ \mathrm{C}$ the Ge surface remains smooth exhibiting no patterns after irradiation (Fig.\ \ref{AFMTemp}(a)). At these conditions the Ge surface is amorphized by ion irradiation and smoothing by surface diffusion and mass redistribution dominates \cite{Madi:2011iy}. The initial roughness, $w_{rms}$, of the virgin Ge surface of 0.7\ nm is reduced to 0.18\ nm. At temperatures higher than $250^\circ \mathrm{C}$ checkerboard patterns appear after irradiation indicating that an additional ion induced instability appears. The structures of the pattern have a rectangular shape and an average size of $\approx150$ nm with an orientation in the $\langle100\rangle$ direction (Fig.\ \ref{AFMTemp}(b)). The size of the structures increases with temperature. A checkerboard pattern with an average size of $\approx260$ nm develops under ion irradiation at $350^\circ \mathrm{C}$ (Fig.\ \ref{AFMTemp}(c)). Again, the pattern is oriented along the $\langle100\rangle$ direction. At $430^\circ \mathrm{C}$ the structure size increases further while the symmetry is changed to an isotropic pattern of pits with diameter of $300 - 1000$ nm. The order is much lower and the structure size distribution much broader than for the checkerboard patterns. Finally, at temperatures above $500^\circ \mathrm{C}$ the surface is again smoothed by ion irradiation, similar to irradiations at $230^\circ \mathrm{C}$ (not shown).  

\begin{figure}
\includegraphics[width=\columnwidth]{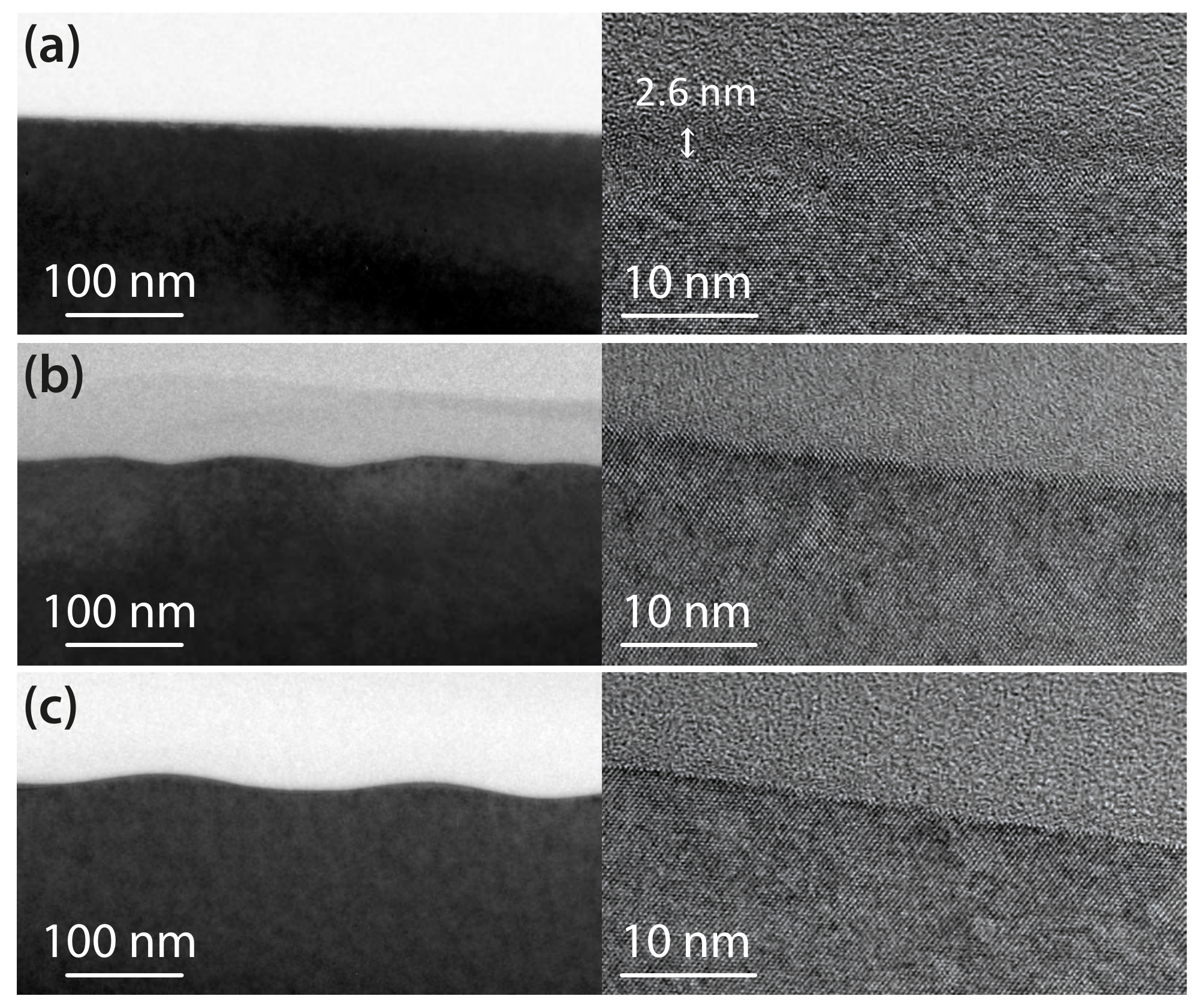}
\caption{\label{TEM}TEM images of Ge surfaces in cross section irradiated at temperatures of (a) $230^\circ$C, (b) $260^\circ$C, and (c) $350^\circ \mathrm{C}$, respectively. In (a) a 2.6 nm thick amorphous layer can be identified on top of the crystalline bulk, whereas in (b) and (c) no amorphous layer is visible.}
\end{figure}

We investigated the microstructure of the patterns in cross section with transmission electron microscopy (TEM). In Fig.\ \ref{TEM} TEM images of Ge (100) surfaces are shown, which were irradiated at temperatures of $230^\circ$C, $260^\circ$C, and $350^\circ$C, respectively (same irradiation conditions as in Fig.\ \ref{AFMTemp}(a)-(c)). At $230^\circ$C the surface is flat and a 2.6 nm thick amorphous Ge layer is visible, in good agreement with the calculated range of 3 nm for 1 keV Ar$^{+}$ ions in Ge \cite{Moller:1988wr}. At $260^\circ$C and higher no amorphous layer is visible. In the high resolution TEM images of the Ge surfaces irradiated at $260^\circ$C and $350^\circ$C flat facets are visible. The angle between the facets and the (100) plane is determined to $8^{\circ}$-$10^{\circ}$. 

The transition from smoothing to roughening by ion irradiation between $230^\circ \mathrm{C}$ and $260^\circ \mathrm{C}$ can be clearly attributed to the temperature at which amorphization by ion irradiation is prevented. Ion induced vacancies and interstitials in the bulk are dynamically annealed above this temperature and only adatoms and surface vacancies remain as defects. Hence, surface patterns result from the kinetics of adatoms and vacancies created by the incident ion beam, however, vacancy kinetics  is expected to dominate, because the number of created vacancies is higher than the number of adatoms by sputtering. In analogy to the ES barrier for adatoms, i.e. the diffusion barrier for an adatom to descent a step edge, a barrier exists for a vacancy to ascend into the next higher terrace. Thus, vacancies are trapped on lower terraces leading to 3D reverse growth of surface structures, i.e. the formation of pyramidal pits (inverse mounds). At higher temperature the mobility of vacancies increases and the nucleation density decreases leading to pits with larger separation and size. If the thermal energy of vacancies is high enough to overcome the ES barrier, the surface will remain smooth which is in agreement with our observation that above $430^\circ \mathrm{C}$ the surface remains flat. 

\begin{figure}
\includegraphics[width=0.9\columnwidth]{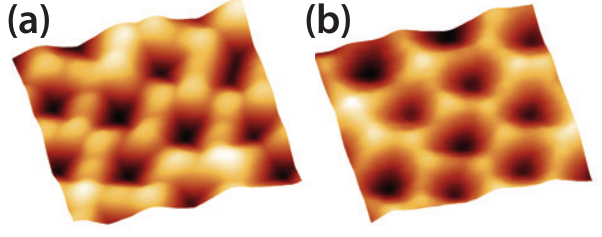}
\caption{\label{AFMzoom}AFM images with size of $500 \times 500\, \mathrm{nm}$ of (a) Ge (100) surface and (b) Ge (111) surface irradiated at temperatures of $280^\circ \mathrm{C}$.}
\end{figure}

According to the terraces-step-kink (TSK) model the evolution of crystalline surfaces is described by atomistic processes on terraces, steps, and kinks \cite{Zhang:1997uo}. Diffusion and attachment of adatoms and vacancies at steps and kinks is expected to be anisotropic on crystalline surfaces. Therefore, the pyramidal pits exhibit facets oriented along the crystalline directions. On Ge (100) the edges of the pits are oriented along the $\langle100\rangle$ crystal directions, whereas on Ge(111) patterns with a six fold symmetry appear (see Fig.\ \ref{AFMzoom}). The emergence of  these edges is attributed to an additional barrier at kink sites, similar to the barrier at step edges \cite{Amar:1999wu}. On the (100) surface the kink ES barrier is responsible for the repulsion of vacancies on the fast diffusing $\langle110\rangle$ step edges. This mechanism has also been proposed for the formation of mounds oriented in the $\langle100\rangle$ directions in homoepitaxy of Ge (100) \cite{Shin:2005ewa}. The energy barrier for the diffusion around corners is expected to be smaller than for crossing step edges. Therefore, the kink ES barrier vanishes already at temperatures where the terrace ES barrier is still active and isotropic pit patterns should appear. This is indeed observed for irradiations at $430^\circ \mathrm{C}$ (Fig.\ \ref{AFMTemp}(d)) where the square symmetry of the pattern disappears and dense round pits are formed. 
 
\begin{figure}
\includegraphics[width=\columnwidth]{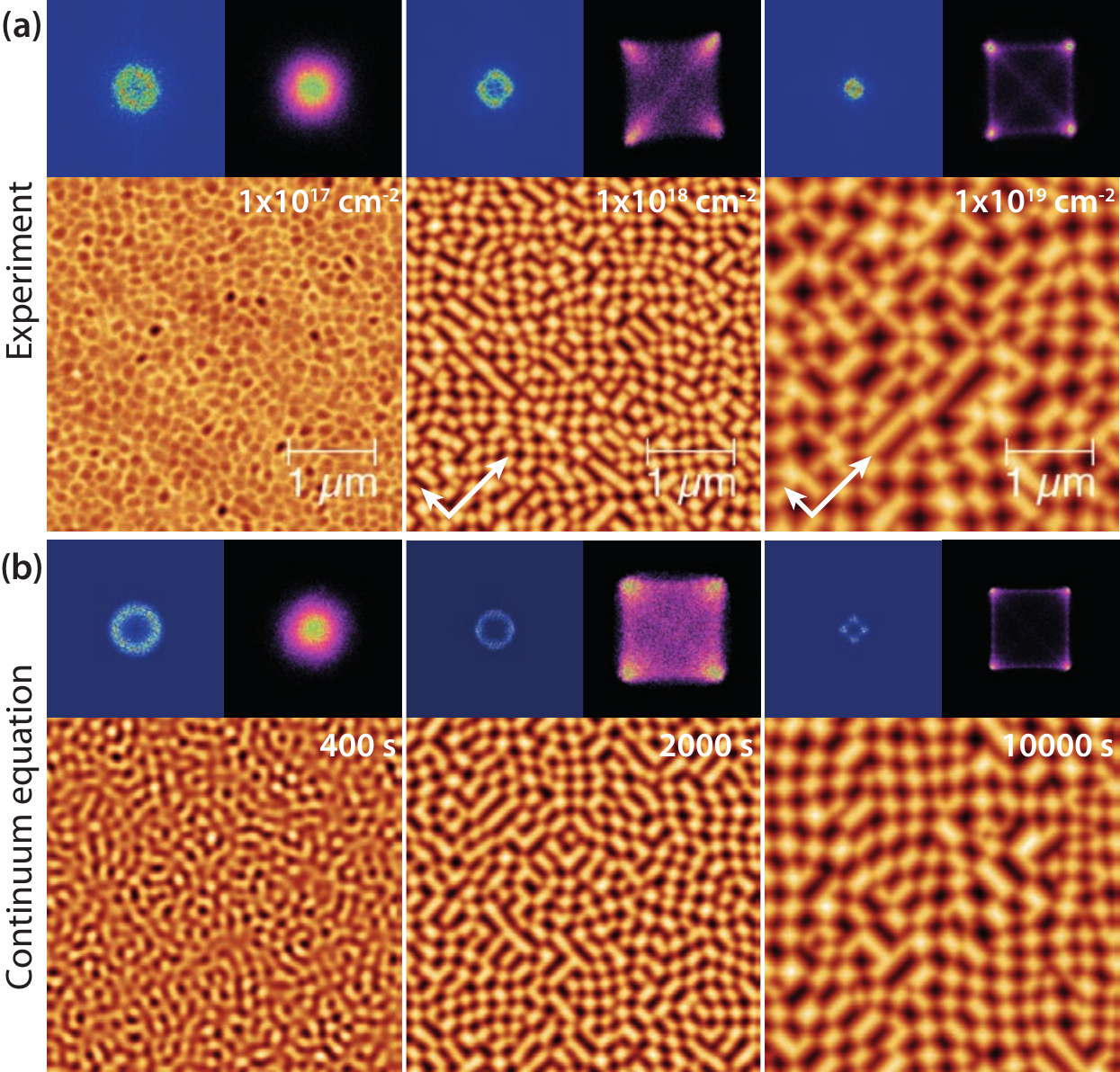}
\caption{\label{AFMfluence}
(a) AFM images of Ge (100) surfaces irradiated at $350^\circ \mathrm{C}$ and fluences of $1\times 10^{17}\mathrm{cm}^{-2}$, $1\times 10^{18}\mathrm{cm}^{-2}$, and $1\times 10^{19}\mathrm{cm}^{-2}$, respectively. The arrows indicate $\langle100\rangle$ directions in the surface.
(b) Snapshots of  the numerical integration of the continuum equation for 400 s, 2000 s, and 10000 s, respectively. The parameters used for the numerical integration are $\epsilon = 1, \kappa=4$, $\sigma=-1$,  and $\delta=25$ on an integration grid of $400 \times 400$ points with $\Delta x= \Delta y = 1$, and $\Delta t = 0.01 s$. Above the height images the 2D FFT (left) and the 2D angle distribution (right) are shown.}
\end{figure}

In order to further elucidate the formation mechanism we investigated the roughening and coarsening behavior of the checkerboard pattern on Ge (100) at $350^\circ \mathrm{C}$. Fig.\ \ref{AFMfluence}(a) shows the evolution of Ge(100) surface as well as the corresponding 2D FFT and 2D angle distribution as a function of ion fluence. The FFT reveals the symmetry and the order of the pattern, whereas from the 2D angle distribution the formation of dominant crystal facets can be deduced. After irradiations with a fluence of $1\times10^{17}\textrm{cm}^{-2}$ a pit pattern with no clear orientation is visible.  The slope distribution is isotropic with a maximum at $0^{\circ}$. At higher fluence of $1\times10^{18}\,\textrm{cm}^{-2}$ the pattern already exhibits an orientation along the $\langle 100\rangle$ crystal direction. The angle distribution reveals a fourfold symmetry with maxima around $9.5^{\circ}$. Finally, at $1\times10^{19}\,\textrm{cm}^{-2}$ the angle distribution has four distinct peaks around $11^{\circ}$. The peaks in the angle distribution are a clear signature for facet formation on the patterned Ge (100) surface. The azimuthal and polar orientation of the facets can be identified with \{106\} ($9.6^{\circ}$) and \{105\} ($10.5^{\circ}$) crystal planes. The angle is determined by the formation of thermodynamic metastable low index planes, which are often observed in Ge homoepitaxy and heteroepitaxy on Si \cite{Teichert:2002uj,Chen:2012ii}. The energy minimization of the (105) facet results from the effective dimerization of the dangling bonds on the two-atom wide (100) terraces separated by $\langle100\rangle$ steps. 



For the description of the temporal evolution of the surface during ion irradiation a continuum equation is derived considering sputtering and mass distribution by the incident ion beam \cite{Bradley:1988uf,Barabasi:ti,Facsko:2004to,Davidovitch:2007cl} and surface diffusion of ion induced vacancies on crystalline surfaces \cite{Siegert:1998gt}. Redistribution and diffusion processes are mass conserving and can therefore be described by surface currents. The temporal evolution of the surface height, $h(x,y,t)$, can be described by the following deterministic partial differential equation:
\begin{equation}
\frac{\partial h}{\partial t} = -v_{0} + \nu \nabla^{2}h - \nabla \bm{ j}_{\mathrm{ion}} - \nabla \bm{ j}_{\mathrm{diff}},  
\end{equation}
where $v_{o}$ is the constant erosion rate of the flat Ge surface, $\nu\nabla^{2}h$ the so-called negative surface tension with $\nu<0$, which takes into account the curvature dependent sputter rate \cite{Bradley:1988uf}, $\bm{ j}_{\mathrm{ion}}$ the surface current resulting from the ballistic mass redistribution induced by momentum transfer from ions to surface atoms \cite{Moseler:2005ka,Davidovitch:2007cl}, and $\bm{ j}_\mathrm{{diff}}$ the surface current due to diffusion \cite{Villain:1991iw}. Mass redistribution by ion impact is also proportional to the surface curvature, $\nabla^{2}h$, with a positive coefficient and dominates the destabilizing sputtering term, $\nu \nabla^{2}h$, at normal and small angles of incidence \cite{Madi:2011iy}. 

On amorphous surfaces, diffusion is isotropic and is described by the Herring-Mullins (HM) surface diffusion current \cite{Mullins:1959vm}. On crystalline surfaces, the diffusive current has to include atomistic surface currents on terraces, across terrace steps, along steps, and across kinks \cite{Michely:2004ua}. On the flat terraces of Ge(100) and Ge(111) an isotropic surface diffusion has to be assumed as well described by the HM surface diffusion, $\bm{j}_\mathrm{HM}$, whereas diffusion across steps is biased by the ES barrier for ascending vacancies, resulting in a net uphill mass current, $\bm{j}_\mathrm{ES}$. The symmetry of the crystalline Ge(100) surface is taken into account by an anisotropic current vector, ${\bm j}_{\mathrm ES}(\bm{m})$, which is a function of the surface slopes, $m_{(x,y)}=\partial_{(x,y)}h$, in $x$ and $y$ direction \cite{Siegert:1994er}:
\begin{equation}
 \bm{ j}_{\mathrm{diff}} = \kappa \nabla (\nabla^{2} h) + \sigma \nabla (\nabla h)^2 + \epsilon \; \left[\begin{array}{c} m_x ( 1- \delta m_x^2)  \\  m_y ( 1 - \delta m_y^2)  \end{array} \right].
\end{equation}   
In Eqn.\ 2 the first term on the rhs is the isotropic HM surface diffusion, $\bm{j}_\mathrm{HM}$, the second term is called ``conserved Kadar-Parisi-Zhang  (KPZ)'' term and has been introduced as a nonlinear current corresponding to the ``non-conserved`` nonlinearity in the KPZ equation ($\lambda/2 (\nabla h)^{2}$) \cite{Villain:1991iw}. This nonlinear current is known to break the up/down symmetry which is prominently seen in the high temperature round pit patterns (Fig.\ \ref{AFMTemp}(d)). However, also the checkerboard patterns in Fig.\ \ref{AFMTemp}(b)-(c) are not symmetric under the transformation $h\rightarrow -h$. For positive values of $\sigma$ the surface evolves to patterns with mound structures, whereas for negative values of $\sigma$ patterns of pit structures appear. Finally, the third term describes the anisotropic ES surface current, $\bm{j}_\mathrm{ES}$, with $\epsilon$ the proportionality factor between the ES current and the surface slope \footnote{The anisotropic ES surface current $\bm{j}_{ES}$ has been introduced with an explicit term coupling the currents in x- and y-direction \cite{Siegert:1998gt}; however, these terms are implicitly already contained in $\nabla \bm{ j}_{{ES}}$ and were therefore omitted for convenience}. For positive $\epsilon$ the ES current is pointing uphill inducing a surface instability \cite{Villain:1991iw}. The anisotropy of the surface current has its origin in the anisotropy of the ES barrier itself as well as in additional currents along step edges due to step edge diffusion \cite{Murty:1996ty,Politi:2000gh}. It is well known that this kind of surface current leads to the formation of facets at angles for which the current becomes zero \cite{Siegert:1998gt,Michely:2004ua}. Close to these points  the surface current is negative (positive) for smaller (larger) angles, leading to an increase (decrease) of the slope. The parameter $\delta$ determines the angle of the facets:  $\theta=\pm\arctan(\sqrt{1/\delta})$. 


The pattern formation at normal incidence is thus dominated by the diffusive current, $\bm{j}_\mathrm{diff}$, in Eqn.\ 1. Curvature dependent sputtering and mass redistribution are effectively smoothing at these conditions and can only reduce the instability in the ES current.
In Fig.\ \ref{AFMfluence}(b) snapshots of the numerical integration of the continuum equation and their corresponding 2D FFTs and 2D angle distribution are shown. After an integration time of 400 s an isotropic pattern forms exhibiting a characteristic periodicity without facets. Facet formation starts around 800 s and at 2000 s facets are already fully developed. The 2D angle distribution reveals distinct peaks at $11^{\circ}$ in diagonal direction of the simulation grid. The 2D FFT shows a circular region already with a slight anisotropy along the k$_{x}$- and k$_{y}$-axes. Finally, at 10000 s very sharp peaks appear in the 2D angle distribution at positions expected from the zeros of the ES surface current. The 2D FFT now also exhibits clear peaks corresponding to the 4-fold symmetry of the pattern. The comparison with the experiments reveals a remarkable agreement. Furthermore, the proposed continuum equation is able to describe the different temperature regimes identified in Fig.\ \ref{AFMTemp} by choosing the proper coefficients \footnote{See Supplemental Material at  URL  for numerical integrations at different temperatures and movies of the height evolution with different coefficients for the low and high temperature case}. 

\begin{figure}
\includegraphics[width=0.9\columnwidth]{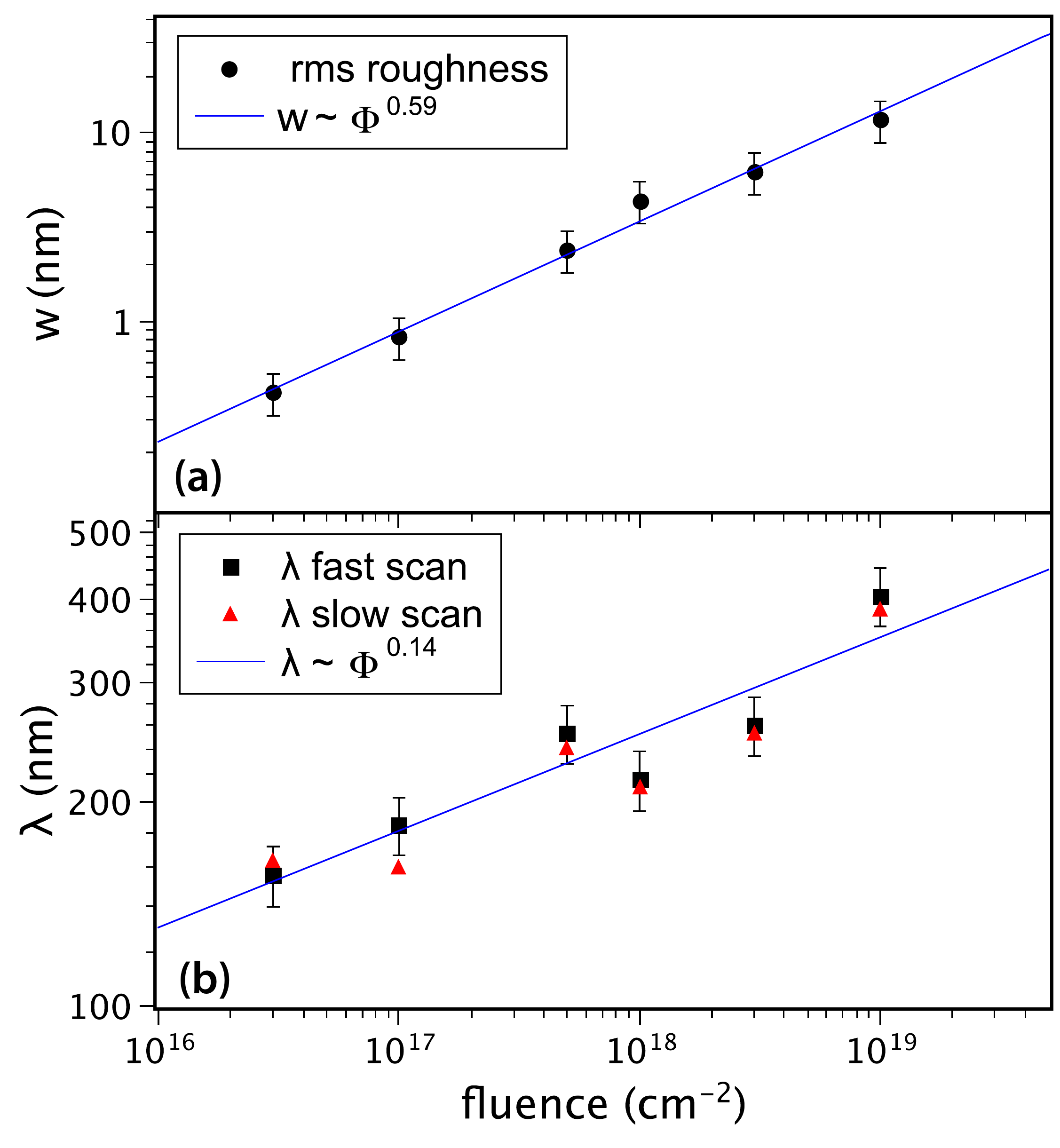}
\caption{\label{RoughCoars}
(a) Roughness and (b) characteristic length of surface patterns on Ge(100) as a function of ion fluence. The characteristic length was extracted from height-height correlation functions along the slow and fast scan direction of the AFM images, respectively. The lines represent power law fits to the data.}
\end{figure}

From the experimental fluence series in Fig.\ \ref{AFMfluence} we determined the temporal evolution of the surface topography. In Fig.\ \ref{RoughCoars}(a) the roughness, $w_{rms}$, is shown as a function of ion fluence. A power law fit to the roughness reveals a growth exponent $\beta=0.59$, which is close to the theoretical value of 0.5 for the so called \emph{statistical growth limit} corresponding to growth by random deposition \cite{Michely:2004ua}. However, due to the existence of an instability larger exponents can be expected \cite{Cuerno:1995ws}. Furthermore, the pattern coarsens with ion fluence, i.e.\ the characteristic length increases with irradiation time, i.e. fluence. In Fig.\ \ref{RoughCoars}(b) the characteristic length of the pattern, $\lambda$, determined from the height-height correlation function of the AFM height images is plotted as a function of ion fluence. The characteristic length of the pattern, $\lambda$, increases according to a power law with a coarsening exponent $1/z=0.14$. Such small coarsening exponents have also been observed for mound coarsening in homoepitaxy \cite{Stroscio:1995ce}. Theoretically a coarsening exponent of 1/4 is predicted for an infinite ES barrier \cite{Amar:1999wu}. Smaller exponents are expected for moderate barriers in step edge diffusion \cite{Amar:1999wu}. The numerical integration of the continuum equation gives a growth exponent of $\beta=0.45$ and a coarsening exponent of $1/z=0.20$, in fair agreement with the experiments.   


Pattern formation by reverse epitaxy is a universal mechanism and can be achieved on many different crystalline materials. We identified the temperature window where patterns of crystalline structures are formed, i.e. the irradiation temperature is (i) above the dynamic recrystallization temperature of the material and (ii) low enough to establish an active ES barrier. At these conditions an excess of vacancies is created  which are partially reflected at terrace steps  inducing an effective uphill mass current. This instability leads to the formation of periodic patterns of inverse pyramids oriented along the crystalline directions of the surface. The faceting of the pyramids results from anisotropic surface currents due to a kink ES barrier. We could thus demonstrate that ion irradiation can induce patterns of facetted crystalline structures by a reverse epitaxy process in analogy to epitaxial growth. The formation of such ordered nanostructured surfaces in reverse epitaxy induced by ion irradiation is considerably easier and less delicate than mound formation in molecular beam epitaxy, where perfect surface preparation and ultrahigh vacuum conditions are crucial. Therefore, this technique could establish as a complementary epitaxial method for the fabrication of high-quality crystalline semiconductor nanostructures.

\begin{acknowledgements}
The authors would like to acknowledge TEM analysis by Arndt M\"ucklich, fruitful discussions with Karl-Heinz Heinig, J\"org Grenzer, Wolfhard M\"oller, and funding from the Deutsche Forschungsgemeinschaft (FOR845).
\end{acknowledgements}

\bibliography{RipplesDots}

\end{document}